\documentclass[prl,twocolumn,superscriptaddress,showpacs]{revtex4-1}
\usepackage{graphicx,amsmath,amssymb,bm}

\newcommand{\kf}{k_{\rm F}}
\newcommand{\kfn}{k_{\rm F}^{\rm n}}
\newcommand{\kfp}{k_{\rm F}^{\rm p}}
\newcommand{\tr}{{\rm Tr}}
\newcommand{\fmi}{\, \text{fm}^{-1}}
\newcommand{\mev}{\, \text{MeV}}

\begin{document}

\title{Chiral three-nucleon forces and pairing in nuclei}

\author{T.\ Lesinski}
\email[E-mail:~]{tlesinsk@uw.edu}
\affiliation{Department of Physics and Institute for Nuclear Theory,
University of Washington, Seattle, WA 98195, USA}
\affiliation{Department of Physics and Astronomy, University of Tennessee,
Knoxville, TN 37996, USA \\
and Physics Division, Oak Ridge National Laboratory, Oak Ridge, TN 37831, USA}
\author{K.\ Hebeler}
\email[E-mail:~]{hebeler.4@osu.edu}
\affiliation{Department of Physics, The Ohio State University,
Columbus, OH 43210, USA}
\affiliation{TRIUMF, 4004 Wesbrook Mall, Vancouver, BC, V6T 2A3, Canada}
\author{T.\ Duguet}
\email[E-mail:~]{thomas.duguet@cea.fr}
\affiliation{CEA, Centre de Saclay, IRFU/Service de Physique Nucl\'eaire,
F-91191 Gif-sur-Yvette, France}
\affiliation{National Superconducting Cyclotron Laboratory
and Department of Physics and Astronomy,
Michigan State University, East Lansing, MI 48824, USA}
\author{A.\ Schwenk}
\email[E-mail:~]{schwenk@physik.tu-darmstadt.de}
\affiliation{ExtreMe Matter Institute EMMI, GSI Helmholtzzentrum f\"ur
Schwerionenforschung GmbH, 64291 Darmstadt, Germany}
\affiliation{Institut f\"ur Kernphysik, Technische Universit\"at
Darmstadt, 64289 Darmstadt, Germany}

\begin{abstract}
We present the first study of pairing in nuclei including
three-nucleon forces. We perform systematic calculations of the
odd-even mass staggering generated using a microscopic pairing
interaction at first order in chiral low-momentum interactions.
Significant repulsive contributions from the leading chiral
three-nucleon forces are found. Two- and three-nucleon interactions
combined account for approximately $70 \%$ of the experimental
pairing gaps, which leaves room for self-energy and induced
interaction effects that are expected to be overall attractive
in nuclei.
\end{abstract}


\maketitle

With the discovery of BCS theory of superconductivity, it was quickly
realized that key nuclear properties, such as the odd-even staggering
of binding energies, or moments of inertia having half the rigid-body
value, were due to the superfluid nature of nuclei~\cite{Bohr58}. In
fact, pairing has become an essential aspect of nuclear structure,
notably for the description of the most proton- and neutron-rich
nuclei~\cite{Doba03}. Although studies of pairing gaps from
internucleon interactions are possible for infinite
matter~\cite{Dean03,Schwenk03,Gezerlis08}, they remain a great
challenge beyond light nuclei.

For medium-mass and heavy nuclei, the method of choice, especially for
systematic global studies, is the nuclear energy density functional
(EDF) approach~\cite{Bender03}. For a single-reference ground state,
pairing is captured through $U(1)$ symmetry breaking and leads to
solving effective Hartree-Fock-Bogoliubov (HFB)~\cite{Ring80} or
Bogoliubov-de Gennes equations. These are solved based on the EDF for
both the single-particle and pairing channels. While current EDF
parameterizations provide a satisfactory description of low-energy
properties of known nuclei, they are empirical in character and lack
predictive power as one enters experimentally unexplored regions. It
is therefore of great interest to construct non-empirical EDFs derived
from microscopic nuclear forces. The development of low-momentum
interactions based on renormalization group (RG)
methods~\cite{Bogner10} opens up this possibility.

Describing pairing within a perturbative expansion around the HFB
state translates into solving a generalized gap equation. This
requires two essential inputs: the normal self-energy that includes
interactions between a single nucleon and the medium, together with
the anomalous self-energy computed from the pairing interaction
kernel. The first-order contribution to the pairing kernel is given
directly by two-nucleon (NN) and three-nucleon (3N) forces (where we
neglect higher-body interactions), while higher-order terms include
induced interactions describing the process of paired particles
interacting via the exchange of medium fluctuations. A fundamental,
yet unresolved, question is to what extent pairing in nuclei is
generated by nuclear forces at first
order~\cite{Duguet08,Lesinski09,Duguet09,Hergert09}, and what is the
role of higher-order
processes~\cite{Terasaki02,Barranco04,Gori05,Pastore08}.

To address this question, we perform a systematic study of pairing
gaps generated using a pairing interaction at first order in chiral
low-momentum interactions~\cite{Bogner10}. Building on
Refs.~\cite{Duguet08,Lesinski09,Duguet09} that explored the
contributions from Coulomb and NN interactions only, we include here
for the first time 3N forces for pairing in nuclei. Higher-order
contributions are left to future works. Since 3N forces play a key role
in neutron-rich nuclei~\cite{3Nnuclei} and matter in neutron
stars~\cite{3Nnstar}, this also presents an important step to
understanding pairing in neutron stars.

Our calculations start from the N$^3$LO NN potential (EM $500 \mev$)
of Ref.~\cite{Entem03}. This is RG-evolved to low-momentum
interactions $V_{{\rm low}\,k}$ using a smooth $n_{\rm exp}=4$
regulator with $\Lambda = 1.8 - 2.8 \fmi$~\cite{smooth,Hebeler07}. In
addition, we include the leading N$^2$LO 3N forces based on chiral EFT
without explicit Deltas~\cite{chiral3N}. They consist of a long-range
2$\pi$-exchange part $V_{c_i}$, an intermediate-range 1$\pi$-exchange part
$V_D$ and a short-range contact interaction $V_E$:
\begin{equation}
\parbox[c]{180pt}{
\includegraphics[scale=0.55,clip=]{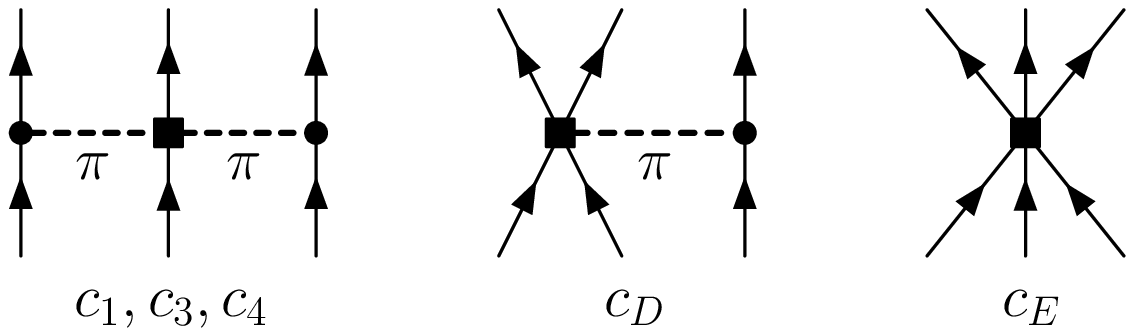}}
\nonumber
\end{equation}
We assume that the $c_i$ coefficients of the long-range 3N parts are
not modified by the RG and use the consistent EM $c_i$'s from the NN
potential of Ref.~\cite{Entem03}. For each cutoff $\Lambda$, we take
the short-range couplings $c_D$ and $c_E$ from the Faddeev and
Faddeev-Yakubovsky fits to the $^3$H binding energy and the $^4$He
matter radius~\cite{Hebeler11}.  This uses a smooth 3N regulator
of the form $\exp[-((p^2 + 3/4 q^2)/\Lambda_{\rm 3NF}^2)^4]$, where
$p$ and $q$ are Jacobi momenta. The 3N cutoff $\Lambda_{\rm 3NF}$ is
allowed to vary independently of the NN cutoff, which probes the
sensitivity to short-range three-body physics. For details and
values of the 3N couplings $c_i, c_D, c_E$, see Ref.~\cite{Hebeler11}.

Based on these 3N forces, we construct an antisymmetrized,
density-dependent two-body interaction $\overline{V}_{\rm 3N}$ by
summing one particle over occupied states in the Fermi sea of
homogeneous nuclear matter, extending the neutron and symmetric matter
calculations of Refs.~\cite{Hebeler10,Hebeler11,Jeremy10} to general
isospin asymmetries:
\begin{equation}
\overline{V}_{\rm 3N}(\kfn,\kfp) = \tr_{\sigma_3,\tau_3} \int
\frac{d{\bf k}_3}{(2\pi)^3} \: \theta(\kf^{\tau_3} - |{\bf k}_3|)
\, \mathcal{A}_{123} \, V_{\rm 3N}\,.
\end{equation}
Here $\mathcal{A}_{123} = 1 - P_{12} - P_{13} - P_{23} + P_{12} P_{23}
+ P_{13} P_{23}$ denotes the three-body antisymmetrizer, where
$P_{ij}$ exchanges spin, isospin and momenta of nucleons $i$ and
$j$. The spin (isospin) projection of the summed particle is denoted
by $\sigma_3$ ($\tau_3$). For neutron matter, only the $c_1$ and $c_3$
parts of 3N forces enter~\cite{Hebeler10}, but with protons present,
all parts contribute at the density-dependent two-body level. In
general, $\overline{V}_{\rm 3N}$ depends on the spin, the isospin, the
relative momenta ${\bf k}, {\bf k}'$ and on the total momentum
${\bf P}$ of the two interacting particles. However, as shown in
Refs.~\cite{Hebeler10,Hebeler11}, the dependence on $\mathbf{P}$ is
very weak and taking ${\bf P}=0$ is a very good approximation.

\begin{figure}[t]
\begin{center}
\includegraphics[scale=0.725,clip=]{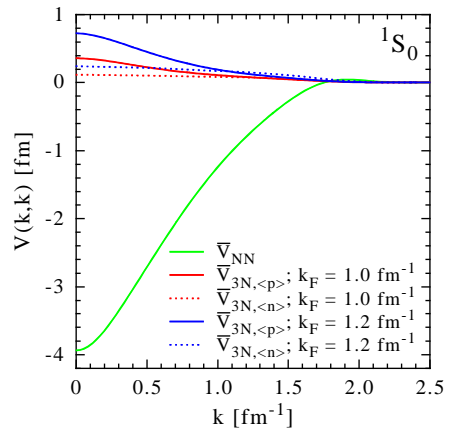}
\end{center}
\vspace*{-3mm}
\caption{Antisymmetrized momentum-space matrix elements for
low-momentum NN and 3N interactions with $\Lambda/\Lambda_{\rm 3NF}
= 2.0/2.0 \fmi$ in the neutron-neutron $^1$S$_0$ channel. The
3N parts are given as density-dependent two-body interactions
$\overline{V}_{\rm 3N}$ for EM $c_i$'s and $c_D, c_E$ from
Ref.~\cite{Hebeler11}. The contributions from summing over
neutrons $\langle {\rm n} \rangle$ and protons $\langle {\rm p}
\rangle$ are shown separately for two Fermi momenta $\kf=1.0,
1.2 \fmi$ in symmetric nuclear matter. The relative momentum $k$
and the $\kf$ dependence is similar for the other 3N fits of
Ref.~\cite{Hebeler11}.\label{fig:matelem}}
\end{figure}

The density-dependent two-body interaction $\overline{V}_{\rm 3N}$
corresponds to the normal-ordered two-body part of 3N
forces~\cite{Hebeler10}. Normal ordering with respect to a superfluid
HFB state leads to a two-body pairing interaction,
where $\overline{V}_{\rm 3N}$ is added with a combinatorial factor $1$
to the antisymmetrized NN interaction $\overline{V}_{\rm NN} =
(1-P_{12}) V_{{\rm low}\,k}$. In this work, we focus on the $^1$S$_0$
partial-wave contribution that dominates isovector
pairing~\cite{Baroni09}, so that the first-order pairing kernel reads
\begin{equation}
\overline{V}_{\rm pairing}^\text{$^1$S$_0$} =
\overline{V}^\text{$^1$S$_0$}_{\rm NN} +
\overline{V}^\text{$^1$S$_0$}_{\rm 3N} \,.
\end{equation}
For practical purposes, it is convenient to separate the contributions
to $\overline{V}_{\rm 3N}$ from the summation over occupied neutron
and proton states, $\overline{V}_{\rm 3N}(\kfn,\kfp) =
\overline{V}_{{\rm 3N}, \langle {\rm n} \rangle}(\kfn) +
\overline{V}_{{\rm 3N}, \langle {\rm p} \rangle}(\kfp)$. The strength
of the two terms depends on the isospin of the two interacting
particles. In Fig.~\ref{fig:matelem}, we compare matrix elements of
the different 3N contributions to the neutron-neutron $^1$S$_0$
pairing kernel for two representative densities of symmetric nuclear
matter. We find that the 3N contributions are both repulsive, so that
pairing will be weaker in nuclei compared to the NN-only level. In
addition, $\overline{V}^\text{$^1$S$_0$}_{{\rm 3N}, \langle {\rm p} \rangle}$
is stronger than $\overline{V}^\text{$^1$S$_0$}_{{\rm 3N}, \langle {\rm n} \rangle}$,
because it involves also isospin $T=1/2$ triples. These 3N effects
can lead to new isospin dependences in pairing gaps.

In order to perform systematic EDF calculations of semi-magic nuclei,
it is convenient to develop an operator representation of
$\overline{V}_{{\rm 3N}, \langle \tau \rangle}(k, k';k^{\tau}_{\rm F})$.
We take a rank-$m$ separable Ansatz of the form
\begin{equation}
\overline{V}^\text{$^1$S$_0$}_{{\rm 3N}, \langle \tau \rangle}(k, k';
k^{\tau}_{\rm F}) = \sum_{\alpha,\beta=1}^m g^{\tau}_\alpha(k) \:
\lambda^{\tau}_{\alpha\beta}(k^{\tau}_{\rm F}) \: g^{\tau}_\beta(k') \,,
\end{equation}
with $m \leqslant 4$. The density dependence is parameterized as a
polynomial in $k^{\tau}_{\rm F}$:
$\lambda^{\tau}_{\alpha\beta}(k^{\tau}_{\rm F}) = \sum_{i=3,4}
\lambda^{\tau}_{\alpha\beta,i} \, (k^{\tau}_{\rm F})^i$. Given our
Ansatz, the functions $g^{\tau}_\alpha(k)$ and the coefficients
$\lambda^{\tau}_{\alpha\beta,i}$ are fitted to the momentum-space
matrix elements $\overline{V}^\text{$^1$S$_0$}_{{\rm 3N}, \langle \tau
\rangle}(k, k'; k^{\tau}_{\rm F})$ for various values of
$k^{\tau}_{\rm F}$ from $0.6$ to $1.6 \fmi$. This describes the matrix
elements to better than $0.01 \, {\rm fm}$. Finally, finite nuclei
calculations use the local approximation
$\lambda_{\alpha\beta}\big(k^{\tau}_{F}({\bf R})\big)$ with
$k^{\tau}_{F}({\bf R}) \equiv \big(3\pi^2 \rho_{\tau}({\bf R})
\big)^{1/3}$. We have checked that an alternate choice of
local-density dependence based on the Campi-Bouyssy prescription for
$k_{\rm F}^\tau({\bf R})$~\cite{Campi78} leads to pairing gaps within
$10 \, {\rm keV}$ of the gaps presented here.

\begin{figure*}[t]
\begin{center}
\includegraphics[scale=0.675,clip=]{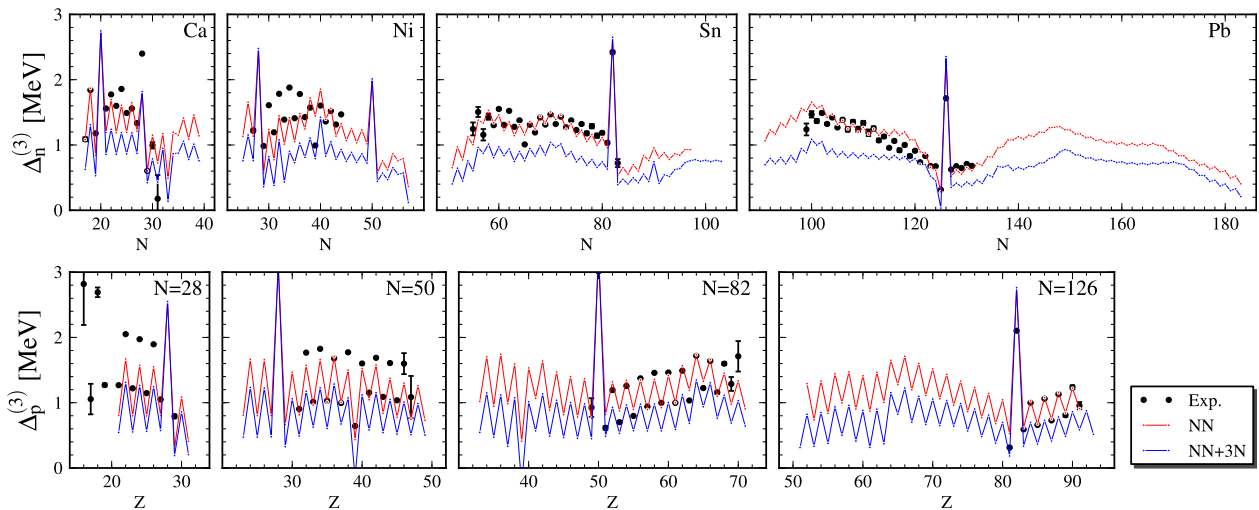}
\end{center}
\vspace*{-5mm}
\caption{Theoretical and experimental neutron/proton three-point mass
differences $\Delta^{(3)}_{\rm n/p}$ along isotopic/isotonic chains
from one-proton to one-neutron drip-lines based on
low-momentum NN and 3N interactions with $\Lambda/\Lambda_{\rm 3NF}
=1.8/2.0 \fmi$. Note that for neutron-rich tin isotopes, the chemical
potential is just below threshold such that the one-neutron drip-line
is located well before the two-neutron drip-line.\label{fig:gaps}}
\end{figure*}

The remaining part of the nuclear EDF, that is due to the normal
self-energy and drives the correlated single-particle motion, is taken
as a semi-empirical Skyrme parameterization. To be as consistent as possible
with the first-order pairing kernel used to compute the anomalous
self-energy, the isoscalar and isovector effective masses of the
Skyrme parameterization have been constrained from Hartree-Fock
results in neutron and symmetric nuclear matter based on low-momentum
NN and 3N interactions~\cite{Hebeler09}. The present results do not
depend significantly on the Skyrme isoscalar effective
mass~\cite{Lesinski08}, as long as its value at saturation density is
within $\approx 0.67\!-\!0.73$ as obtained from the microscopic Hartree-Fock
calculations.

We use the BSLHFB code~\cite{Lesinski07} that solves the
HFB equations in a spherical box of $24
\, {\rm fm}$ radius, with a mesh size of $0.3 \, {\rm fm}$. The
single-particle wave functions are expanded on a basis of spherical
Bessel functions $j_\ell(kr)$ with a momentum cutoff $k_\text{cut} =
4.0 \fmi$, allowing the description of single-particle states up to
energies of about $300 \mev$ and ensuring convergence of the pairing
gaps to a fraction of a keV.

In this Letter, we focus on the odd-even staggering of nuclear masses
that provides a measure of the pairing gap associated with the lack of
binding of an odd isotope/isotone relative to its even neighbors. The
three-point mass differences are computed in the exact same way from
experimental data and EDF calculations. To do so, odd-even nuclei are
computed through the self-consistent blocking procedure performed
within the filling approximation~\cite{PerezMartin08,Duguet09}. In
Fig.~\ref{fig:gaps}, we present the central results for theoretical
and experimental neutron/proton three-point mass differences
$\Delta^{(3)}_{\rm n/p}$ along several semi-magic isotopic/isotonic
chains. Results obtained with and without 3N contributions to the
first-order pairing interaction kernel are compared.

The main result obtained with NN only is that theoretical neutron and
proton pairing gaps computed at lowest order are close to experimental
ones for a large set of semi-magic spherical nuclei, although
experimental gaps are underestimated in the lightest systems. The
addition of the first-order 3N contribution then lowers pairing gaps
systematically by about $30\%$. This is in line with the repulsive
$\overline{V}_{\rm 3N}$ in the $^1$S$_0$ channel (Fig.~\ref{fig:matelem}).

Although the impact of the 3N contribution is generally smooth as a
function of $(N,Z)$ and insensitive to the structure of the particular
nucleus under consideration, it displays a slight isovector
trend. This is seen, e.g., in the lesser reduction of gaps in lead
isotopes with $N>140$ ($-30\%$ average relative shift) than $N<140$
($-35\%$). Similarly, in the $N=126$ isotones we find $-43\%$ for
$Z<70$ and $-35\%$ for $Z>70$.
This effect may be explained by the fact that the interaction involving
three neutrons (protons) is less repulsive than the interaction of two
neutrons (protons) with a proton (neutron). The resulting density-dependent
pairing interaction thus suppresses the neutron (proton) gap
less (more) strongly in a neutron-rich, proton-poor nucleus than in a
symmetric one.

\begin{figure}[t]
\begin{center}
\includegraphics[scale=0.575,clip=]{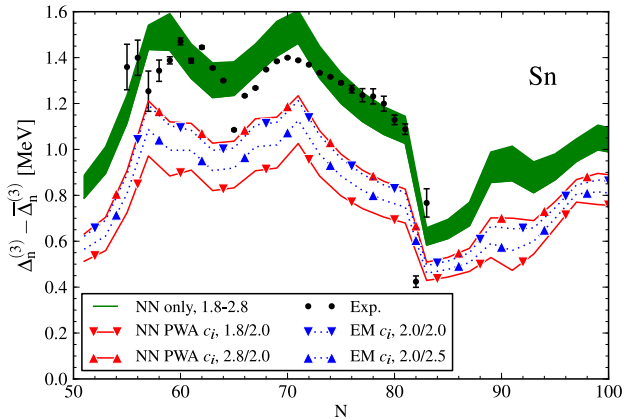}
\end{center}
\vspace*{-3mm}
\caption{Cutoff dependence of the three-point mass difference
$\Delta^{(3)}_{\rm n}$ along the tin isotopic chain, with the
oscillating part $\overline{\Delta}^{(3)}_{\rm n}$
subtracted. Results are shown for the NN cutoff range
$\Lambda=1.8-2.8 \fmi$ (NN-only gaps are the green/solid
band) and including 3N forces with different EM/PWA $c_i$ couplings
and different NN/3N cutoffs $\Lambda/\Lambda_{\rm 3NF}$
(lines/symbols).\label{fig:cutoffdep}}
\end{figure}

Next, we study the cutoff dependence of the pairing gaps in
Fig.~\ref{fig:cutoffdep}. This provides an estimate of the theoretical
uncertainties due to short-range higher-order NN and many-body
interactions as well as due to an incomplete many-body treatment. To
make the comparison clearer, we subtract the oscillating part from the
three-point mass differences and consider
$\Delta^{(3)}-\overline{\Delta}^{(3)}$, where
$\overline{\Delta}^{(3)}$ accounts for that oscillation and is
obtained by treating the odd nuclei as if they had the same structure
as the even ones~\cite{Duguet02}. For the large NN cutoff range
considered, the pairing gaps at the NN-only level vary by $\approx
100-200 \, {\rm keV}$, and even smaller for less smooth cutoffs. When
3N forces are included, the theoretical uncertainties are of similar
size, $\approx 100-250 \, {\rm keV}$, as indicated by the range of
dotted and solid lines in Fig.~\ref{fig:cutoffdep}. This range
includes estimates from neglected shorter-range many-body interactions
that are probed when varying the 3N cutoff ($\Lambda_{\rm 3NF} =
2.0-2.5 \fmi$ for the dotted lines in Fig.~\ref{fig:cutoffdep}) and
from the uncertainties in the long-range $c_i$ couplings (dotted
versus solid lines, where in addition to the consistent EM $c_i$'s, we
consider the central $c_i$ values obtained from the NN partial wave
analysis (PWA)~\cite{PWAci}). For all 3N forces, the short-range
couplings $c_D, c_E$ are taken from Ref.~\cite{Hebeler11}. Similar
cutoff dependences are found for the other semi-magic chains.

In summary, we have carried out the first study of pairing in nuclei
with 3N forces. Our results show that (i) it is essential to include
3N contributions to the pairing interaction for a quantitative
description of nuclear pairing gaps, (ii) our first-order low-momentum
results leave about $30 \%$ room for contributions from higher orders,
e.g., from the coupling to (collective) density, spin and isospin
fluctuations, consistent with induced interactions being overall
attractive in nuclei~\cite{Terasaki02,Barranco04,Gori05,Pastore08},
(iii) in the next steps, the normal self-energy and higher-order
contributions to the pairing kernel must be computed consistently
based on low-momentum NN and 3N interactions. Work in these
directions is in progress. Finally, there are indications from
phenomenological studies that the quality of the overall agreement
between theory and experiment is different for spherical and deformed
nuclei~\cite{Bertsch08,Yamagami08}. It is therefore of great interest
to apply the developed non-empirical pairing energy functional to
deformed nuclei.

\begin{acknowledgments}
This work was supported in part by the DOE under Grants
DE-FG02-07ER41529 and DE-FG02-00ER41132, the UNEDF SciDAC
Collaboration under DOE Grant DE-FC02-07ER41457, the NSF under Grants
PHY-0835543 and PHY-1002478, by NSERC, the Helmholtz Alliance Program
of the Helmholtz Association, Contract HA216/EMMI ``Extremes of
Density and Temperature: Cosmic Matter in the Laboratory'', and the
DFG through grant SFB 634. TRIUMF receives funding via a
contribution through the NRC Canada.
\end{acknowledgments}


\begin{thebibliography}{99}
\bibitem{Bohr58} A.\ Bohr, B.\ R.\ Mottelson and D.\ Pines,
Phys. Rev. \textbf{110}, 936 (1958).

\bibitem{Doba03} J.\ Dobaczewski and W.\ Nazarewicz,
Prog.\ Theor.\ Phys.\ Suppl.\ \textbf{146}, 70 (2003).

\bibitem{Dean03} D.\ J.\ Dean and M.\ Hjorth-Jensen,
Rev.\ Mod.\ Phys.\ \textbf{75}, 607 (2003).

\bibitem{Schwenk03} A.\ Schwenk, B.\ Friman and G.\ E.\ Brown,
Nucl.\ Phys.\  A \textbf{713}, 191 (2003).

\bibitem{Gezerlis08} A.\ Gezerlis and J.\ Carlson,
Phys.\ Rev.\ C \textbf{77}, 032801 (2008).

\bibitem{Bender03} M.\ Bender, P.-H.\ Heenen and P.-G.\ Reinhard,
Rev.\ Mod.\ Phys.\ \textbf{75}, 121 (2003).

\bibitem{Ring80} P.\ Ring and P.\ Schuck,
\textit{The Nuclear Many-Body Problem} (Springer, 1980).

\bibitem{Bogner10} S.\ K.\ Bogner, R.\ J.\ Furnstahl and
A.\ Schwenk, Prog.\ Part.\ Nucl.\ Phys. {\bf 65}, 94 (2010).

\bibitem{Duguet08} T.\ Duguet and T.\ Lesinski,
Eur.\ Phys.\ J.\ ST \textbf{156}, 207 (2008).

\bibitem{Lesinski09} T.\ Lesinski {\it et al.},
Eur.\ Phys.\ J.\ A \textbf{40} 121 (2009).

\bibitem{Duguet09} T.\ Duguet and T.\ Lesinski,
AIP Conf.\ Proc.\ \textbf{1165}, 243 (2009).

\bibitem{Hergert09} H.\ Hergert and R.\ Roth,
Phys.\ Rev.\ C \textbf{80}, 024312 (2009).

\bibitem{Terasaki02} J.\ Terasaki {\it et al.},
Nucl.\ Phys.\ A \textbf{697}, 127 (2002).

\bibitem{Barranco04} F.\ Barranco {\it et al.},
Eur.\ Phys.\ J.\ A \textbf{21}, 57 (2004).

\bibitem{Gori05} G.\ Gori {\it et al.},
Phys.\ Rev.\ C \textbf{72}, 1 (2005).

\bibitem{Pastore08} A.\ Pastore {\it et al.},
Phys.\ Rev.\ C \textbf{78}, 024315 (2008).

\bibitem{3Nnuclei} T.\ Otsuka {\it et al.},
Phys.\ Rev.\ Lett.\ {\bf 105}, 032501 (2010);
J.\ D.\ Holt {\it et al.}, arXiv:1009.5984.

\bibitem{3Nnstar} K.\ Hebeler  {\it et al.},
Phys.\ Rev.\ Lett.\ {\bf 105}, 161102 (2010).

\bibitem{Entem03} D.\ R.\ Entem and R.\ Machleidt,
Phys.\ Rev.\ C \textbf{68}, 041001(R) (2003).

\bibitem{smooth} S.\ K.\ Bogner {\it et al.},
Nucl.\ Phys.\ A \textbf{784}, 79 (2007).

\bibitem{Hebeler07} K.\ Hebeler, A.\ Schwenk and B.\ Friman,
Phys.\ Lett.\ B \textbf{648}, 176 (2007).

\bibitem{chiral3N} U.\ van Kolck,
Phys.\ Rev.\ C {\bf 49}, 2932 (1994);
E.\ Epelbaum {\it et al.},
Phys.\ Rev.\ C {\bf 66}, 064001 (2002).

\bibitem{Hebeler11} K.\ Hebeler {\it et al.},
Phys.\ Rev.\ C \textbf{83}, 031301(R) (2011).

\bibitem{Jeremy10} J.\ W.\ Holt, N.\ Kaiser and W.\ Weise,
Phys.\ Rev.\ C \textbf{81}, 024002 (2010).

\bibitem{Hebeler10} K.\ Hebeler and A.\ Schwenk,
Phys.\ Rev.\ C \textbf{82}, 014314 (2010).

\bibitem{Baroni09} S.\ Baroni, A.\ O.\ Macchiavelli and A.\ Schwenk,
Phys.\ Rev.\ C \textbf{81}, 064308 (2010).

\bibitem{Campi78} X.\ Campi and A.\ Bouyssy,
Phys.\ Lett.\ B \textbf{73}, 263 (1978).

\bibitem{Hebeler09} K.\ Hebeler {\it et al.},
Phys.\ Rev.\ C \textbf{80}, 044321 (2009).

\bibitem{Lesinski08} T.\ Lesinski, T.\ Duguet, K.\ Bennaceur and
J.\ Meyer, unpublished.

\bibitem{Lesinski07} T.\ Lesinski,
Ph.D.\ Thesis (Universit\'e Lyon 1, 2008).

\bibitem{PerezMartin08} S.\ Perez-Martin and L.\ M.\ Robledo,
Phys.\ Rev.\ C \textbf{78}, 014304 (2008).

\bibitem{Duguet02} T.\ Duguet {\it et al.},
Phys.\ Rev.\ C \textbf{65}, 014311 (2002).

\bibitem{PWAci} M.\ C.\ M.\ Rentmeester, R.\ G.\ E.\ Timmermans and
J.\ J.\ de Swart, Phys.\ Rev.\ C \textbf{67}, 044001 (2003).

\bibitem{Bertsch08} G.\ F.\ Bertsch {\it et al.},
Phys.\ Rev.\ C \textbf{79}, 034306 (2009).

\bibitem{Yamagami08} M.\ Yamagami, Y.\ R.\ Shimizu and
T.\ Nakatsukasa, Phys.\ Rev.\ C \textbf{80}, 064301 (2009).
\end{thebibliography}
\end{document}